\documentclass[preprint,secnumaRabic,amssymb,nobibnotes,aps,pra]{revtex4-2}

\usepackage{xcolor}
\usepackage[normalem]{ulem}
\usepackage{graphicx}
\usepackage{dcolumn}
\usepackage{bm}
\usepackage{amsmath}

\begin{document}

\title{
Closed Loop Quantum Interferometry for\\ 
Phase-Resolved Rydberg Atom Field Sensing
}

\author{Samuel Berweger$^1$}
\author{Alexandra B. Artusio-Glimpse$^1$}
\author{Andrew P. Rotunno$^1$}
\author{Nikunjkumar Prajapati$^1$}
\author{Joseph D. Christesen$^2$}
\author{Kaitlin R. Moore$^3$}
\author{Matthew T. Simons$^1$}
\author{Christopher L. Holloway$^1$}
\affiliation{$^1$National Institute of Standards and Technology, Boulder, CO, 80305}
\affiliation{$^2$SRI International, Boulder, CO, 80302}
\affiliation{$^3$SRI International, Princeton, NJ, 08540}

\begin{abstract}
Although Rydberg atom-based electric field sensing provides key advantages over traditional antenna-based detection, it remains limited by the need for a local oscillator (LO) for low-field and phase resolved detection.
In this work, we demonstrate that closed-loop quantum interferometric schemes can be used to generate a system-internal reference that can directly replace an external LO for Rydberg field sensing.
We reveal that this quantum-interferometrically defined internal reference phase and frequency can be used analogously to a traditional LO for atom-based down-mixing to an intermediate frequency for lock-in phase detection.
We demonstrate that this LO-equivalent functionality provides analogous benefits to an LO, including full 360$^\circ$ phase resolution as well as improved sensitivity.
The general applicability of this approach is confirmed by demodulating a four phase-state signal broadcast on the atoms.
Our approach opens up new sensing schemes and \textcolor{black}{although the present implementation still uses an auxiliary RF field, it} provides a clear path towards all-optical Rydberg atom sensing implementations.
\end{abstract}

\maketitle

\section{Introduction}
Rydberg atom-based field sensing is an emerging technology that uses resonant transitions between excited states at high principal quantum numbers, $n$, to detect radio frequency (RF) electric (E) fields \cite{Aly2022,sedlacek2012,fan2015}.
This technology has the potential to replace traditional wavelength-scaling antenna architectures with compact atomic vapor cells \cite{cox2018} that use an optical readout of the atomic response.
However, similar to traditional RF demodulation schemes, phase-sensitive detection often requires a local reference field.
For the case of Rydberg atoms, an additional local oscillator (LO) field can be applied, which will be mixed to an intermediate frequency (IF) by the atoms themselves (``Rydberg mixer``) \cite{simons2019,jing2020}.
This Rydberg mixer provides benefits including improved sensitivity \cite{gordon2019,jing2020}, frequency selectivity \cite{meyer2021,liu2022}, and phase sensitivity \cite{meyer2018} that allows angle-of-arrival \cite{Amy2021} detection and demodulation of phase-modulated communication signals \cite{holloway2019}.
However, a Rydberg mixer nevertheless requires an additional LO RF field radiating the atoms with a frequency within a few MHz of the measured field and a matched phase front, which can be difficult to achieve and is undesirable in many applications.

One possible way to eliminate the need for an externally applied LO is to use a closed loop scheme \cite{morigi2002,shylla2018}.
These schemes exploit the quantum mechanical interference across a set of driven transitions between discrete states that form a closed loop.
These can be used to mutually reference the phases of fields across large frequency ranges \cite{huss2004}.
Any transition between states in this loop will simultaneously occur in both directions, where interference between the two paths results in a transition probability that depends on the relative phases of all fields involved \cite{kajarischroder2007}.
Such approaches have typically been applied to atomic ground-state transitions \cite{huss2004}, \textcolor{black}{but more recently Rydberg atom loop schemes have emerged as an attractive means for phase transfer between optical and microwave photons for quantum communications applications \cite{han2018,kumar2023}}. 
Closed-loop schemes can be complex because of orbital angular momentum selection rules that require at least four transitions, and in this respect a proposed scheme for Rydberg sensing that requires four RF fields is impractical \cite{shylla2018}.
A recent experimental implementation that drives two degenerate RF transitions succeeded at eliminating the need for additional RF fields but was unable to achieve the full 360$^\circ$ phase resolution necessary for modern digital modulation schemes such as phase key shifting \cite{anderson2022}.

In this work, we demonstrate the general applicability of closed loop schemes using Rydberg states for phase-sensitive field sensing applications and show how they can directly produce LO-equivalent functionality.
We implement a quantum interferometric loop scheme for RF sensing where we leverage the versatility of such an approach by using a loop where the four transitions are comprised of two optical and two non-degenerate RF frequencies. 
We show that this scheme provides full 360$^\circ$ phase resolution on both RF fields.
Furthermore, we clearly demonstrate that a closed loop scheme establishes an LO-free quantum coherent reference frequency and phase that can be exploited analogously to the LO used in established Rydberg mixer measurements \cite{simons2019}.
Using this quantum reference we perform LO-free demodulation of a quadrature phase shift key (QPSK)-equivalent four-phase state signal at a symbol rate of 800 Hz.
We reveal that the sensitivity relative to a traditional LO-based Rydberg mixer is reduced by as little as a factor of 5, and we expect significantly improved sensitivity and bandwidth using optimally chosen states \cite{chopinaud2021}. 
\textcolor{black}{Although the present implementation still requires an auxiliary RF field}, a notable feature of this scheme is the possibility of an RF-free all-optical implementation that closes the loop using three phase-locked optical fields to measure a fourth RF field. 

\begin{figure*}[!htbp]
\centering
\includegraphics[width=.95\linewidth]{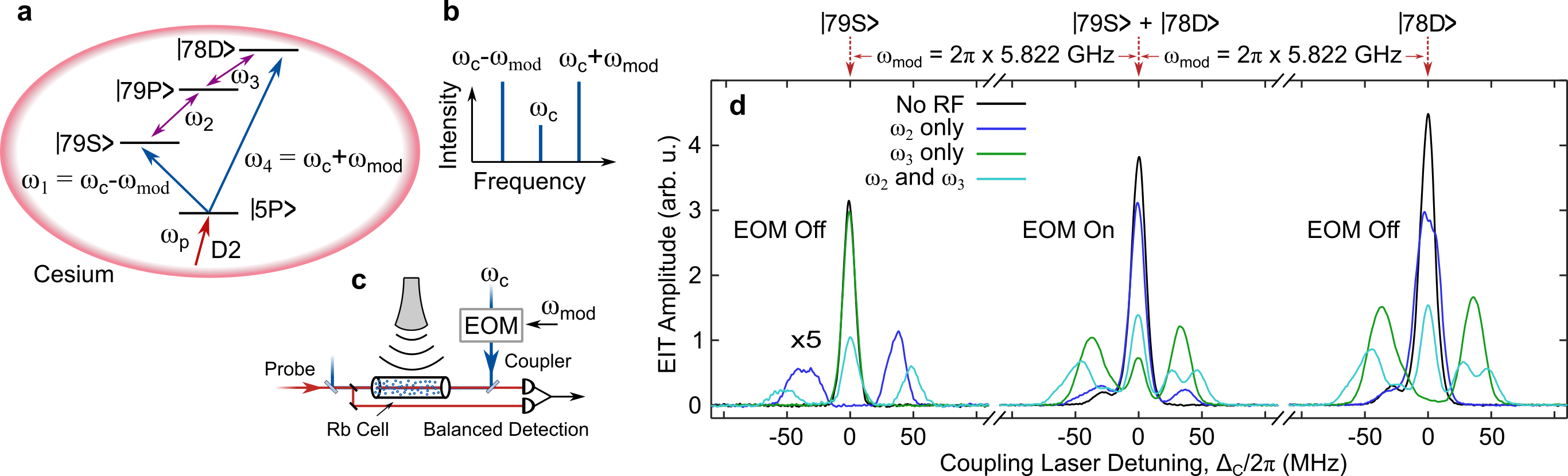}
\caption{\textbf{Experimental Details.} (a) Schematic of the EIT ladder and Rydberg states used for the quantum interference scheme, as well as (b) the phase-modulating electro-optic modulator (EOM) used for coupling laser sideband generation. 
(c) Schematic of the experimental setup with counterpropagating probe and coupling beams.
(d) EIT spectra of the Rydberg states used with the two RF fields turned on and off as indicated.
}
\label{setup}
\end{figure*}

\section{Experimental}

A schematic of the closed-loop scheme used for our experiment is shown in Fig.~\ref{setup}(a). 
The loop is comprised of four fields, $E_i, i=1,\ldots4$, each with corresponding frequencies $\omega_i$, Rabi frequencies $\Omega_i$, and phases $\phi_i$.
The probe field on the D2 transition first couples the $^{85}$Rb 5S$_{1/2}$ ground state to the 5P$_{3/2}$ state. 
An electro-optic modulator (EOM) driven by an external phase-stable signal at a frequency $\omega_{mod}=2\pi\times$~5.822~GHz generates sidebands on the coupling field (Fig.~\ref{setup}(b)) at $\omega_1$~=~$\omega_c-\omega_{mod}$ and $\omega_4$~=~$\omega_c$~+~$\omega_{mod}$ that then couple to the 79S$_{1/2}$ and 78D$_{5/2}$ states, respectively.
These states are then linked through the 79P$_{3/2}$ state via two RF-frequency transitions at $\omega_2=2\pi\times$~7.292~GHz (SP transition) and $\omega_3=2\pi\times$~4.352~GHz (DP transition).
The phases and frequencies of this loop state arrangement are related by $\omega_1+\omega_2+\omega_3=\omega_4$ and $\phi_1+\phi_2+\phi_3=\phi_4$.
Using the two EOM-generated sidebands (Fig.~\ref{setup}(b)), these relationships reduce to $\omega_2+\omega_3-2\omega_{mod}=0$ and $\phi_2+\phi_3 -2\phi_{mod}=0$.
One notable benefit of this scheme is that any frequency or phase noise and/or drift in the coupling laser cancels out, with the only remaining dependence on our phase-locked RF fields.
This set of states is chosen based on the narrow bandwidth of our EOM around 5.8~GHz, but this approach is generally applicable and we have verified that it works for other closed-loop state manifolds.

Our setup uses the established electromagnetically-induced transparency (EIT) ladder approach to excite and probe the high lying Rydberg states \cite{mohapatra2008}. 
Shown in Fig.~\ref{setup}(c) is a schematic of the experimental setup, consisting of counterpropagating coupling and probe lasers that are spatially overlapped in a rubidiuim vapor cell and the RF fields are broadcast onto the cell using a standard gain horn antenna.
The EIT-induced change in probe transmission is detected using balanced photodetection.
The EOM is inserted into the coupling beam path to generate sidebands on the coupling laser frequency  $\omega_c$, located at $\omega_1$ and $\omega_4$ as described above.

\section{Results}

We begin by examining the EIT spectra of our system of states.
Shown in Fig.~\ref{setup}(d) are a set of spectra of the bare 79S and 78D states (left and right, respectively) with the EOM turned off, where the 79S state EIT amplitude is approximately 5 times weaker than the 78D. 
As we probe the bare states, we see Autler-Townes (AT) splitting when the RF fields corresponding to the adjacent transitions are turned on ($\omega_2$ for 79S and $\omega_3$ for 78D, Fig.~\ref{setup}(d) left and right plots, respectively), where we ensure that the Rabi frequencies are equivalent: $\Omega_2=\Omega_3=2\pi\times$~80~MHz.
In contrast, when only the non-adjacent RF fields are applied ($\omega_2$ for 79S and $\omega_3$ for 78D), there is little effect on the EIT spectra; though a limited effect on the 78D state is seen due to the nearby transition to 76F$_{7/2}$. 
\textcolor{black}{This nearby transition is also likely responsible for the asymmetry seen in the AT doublet}.
When both RF fields are applied, we see two key effects: the AT-splitting further increases and the central EIT peak reappears.
The reappearance of the EIT peak is due to a two-photon Raman transition that we described previously \cite{berweger2022}, which here resonantly links the 79S and 78D states.

We set up our interferometric loop by turning on the EOM and setting the coupling laser frequency halfway between the 79S and 78D states, i.e., the sidebands are on-resonance with these transitions. 
As a result, the sideband-generated 79S and 78D EIT peaks are spectrally superimposed when no RF is applied in Fig.\ref{setup}(d) (center).
With applied $\omega_2$ or $\omega_3$ RF fields, the corresponding constituent 79S and 78D peaks undergo AT splitting and the other peak remains unchanged as a residual central EIT peak.
This superposition peak is dominated by the significantly stronger contribution of the 78D transition.
When both RF fields are applied -- thus completing the interferometric loop -- we see a superposition of the AT-doublets from the 79S and 78D states, as well as a central EIT peak that is due to the two-photon Raman transition.
It is this configuration with the superimposed AT doublets and the two-photon Raman peak that we will use to examine the effect of RF phase.

\begin{figure*}[!htb]
\centering
\includegraphics[width=.65\linewidth]{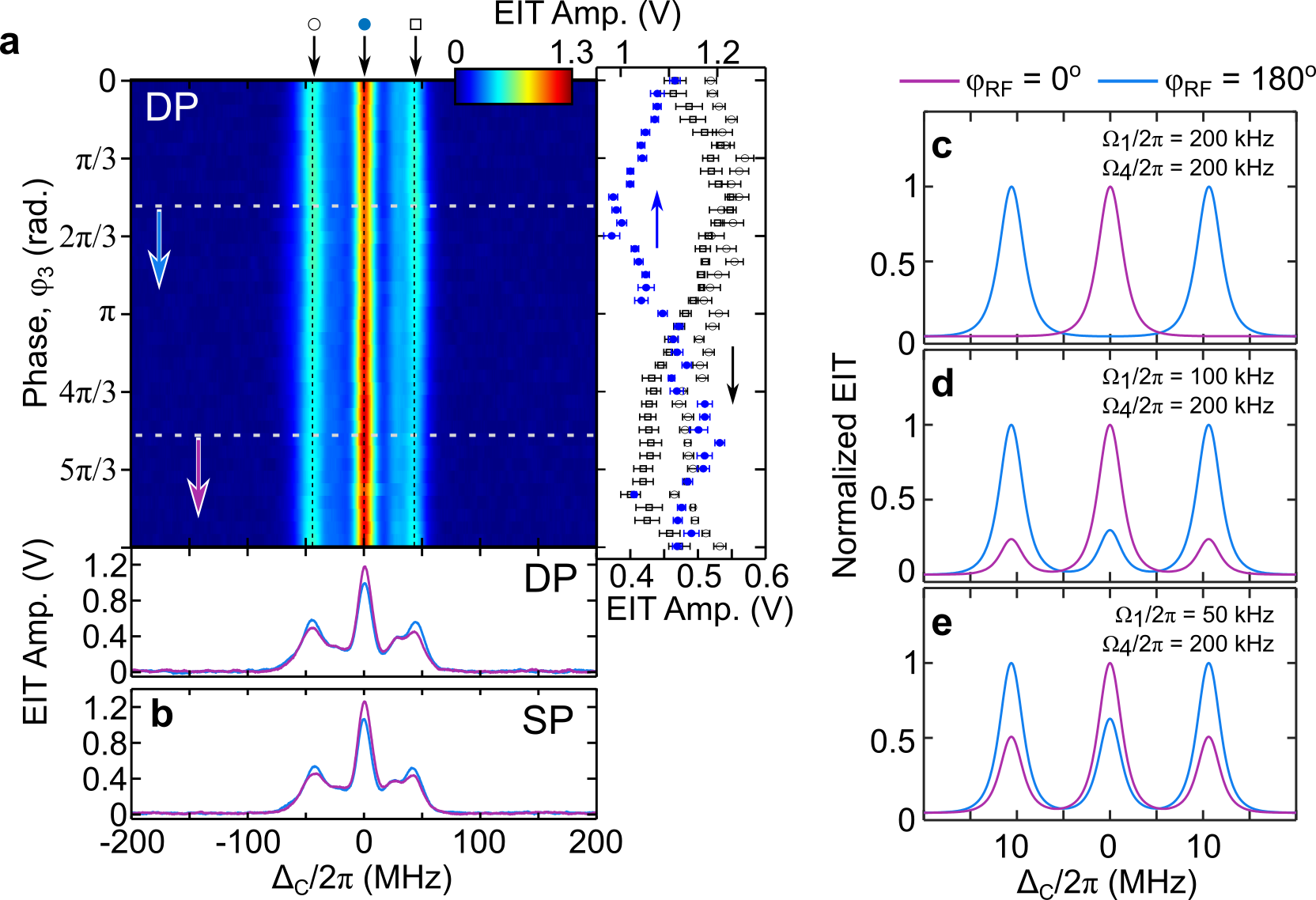}
\caption{\textbf{Phase Sensitivity.}
Demonstration of phase sensitivity on the DP transition (a) while the phase of the other field is held constant. 
The false-color plot shows the evolution of the EIT amplitude as a function of the corresponding RF phase ($\phi_3)$.
The right panel shows the line cuts along each of the prominent EIT peaks as a function of phase and the bottom panel shows the EIT spectra taken along the dashed lines in the false color plot indicated by the colored arrows.
Spectra showing the equivalent maxima in EIT amplitude as a function of the RF phase applied to the SP transition ($\phi_2$) is shown in (b).
Modeled results in (c)--(e) show the evolution of the phase modulation depth as a function of the difference in coupling laser Rabi frequencies. 
}
\label{PhaseSweep}
\end{figure*}

We begin our examination of the RF phase with the effect of the phase of $\omega_3$ ($\phi_3$).
Shown in Fig.~\ref{PhaseSweep}(a) is a false color plot of the superposition EIT peak amplitude as $\phi_3$ is swept over 360$^\circ$.
Phase-dependent (vertical) line cuts taken from the central (blue circles) and side peaks (open black circles and squares), shown on the right reveal a clear oscillation in the amplitude with a depth of around 20$\%$ and a periodicity of 360$^\circ$.
\textcolor{black}{The error bars in the phase-dependent line cuts are calculated from the standard deviation of 10 measurements taken in sequence, and primarily reflect the noise of the probe laser while larger variations arise from fluctuations in our balanced detection due to thermal variations and drift.}
As confirmed in spectral (horizontal) line cuts taken along the white dashed lines and shown on the bottom, the central and AT-split side peaks oscillate out of phase.

Since the accumulated phase of our quantum interferometric loop is due to all fields involved, our measurement is sensitive to changes in phase of any of the fields. 
Shown in Fig.~\ref{PhaseSweep}(b) is the response of the EIT signal to the phase of $\omega_2$ ($\phi_2$), where a comparable 20$\%$ modulation of the peak is also seen.

While our modulation depth is only around 20$\%$, this is comparable to the relative amplitudes of the 79S and 78D EIT peaks that we are superimposing for our measurements.
The optical field strengths of our EOM-generated sidebands are locked at the same value, so any differences in peak amplitudes are \textcolor{black}{expected to be} due to the transition dipole moments of the coupling laser transitions.
\textcolor{black}{However, with transition dipole moments $\mu_4 \approx 2\mu_1$ we similarly expect a factor of two difference in the EIT amplitudes.
While we see EIT amplitudes differing by a factor of 2 for $n<60$ -- in good agreement in our modeled EIT amplitudes -- we see a large difference here for $n \approx 78$.
The origin of this discrepancy is unclear, but we expect that it relates to the increasingly higher density of nearby high angular momentum states at large $n$.}

We theoretically investigate the effect of the optical Rabi frequencies $\Omega_1$ and $\Omega_4$ on the phase-dependent modulation depths.
Shown in Fig.~\ref{PhaseSweep}(c)-(e) are the modeled \cite{holloway17,berweger2022} EIT amplitudes at in-phase and out-of-phase conditions with $\Omega_1$ and $\Omega_4$ as indicated.
Here we see that a full modulation depth can be achieved if the Rabi frequencies \textcolor{black}{-- and thus EIT amplitudes --} of the two fields are the same, which decreases to 42$\%$ for $\Omega_1=4 \times \Omega_4$.
We note that although quantitative changes in the modulation depth depends on the effective Rabi frequency, larger differences between $\Omega_1$ and $\Omega_4$ always lead to reduced contrast. 
These models also confirm the experimental observation noted above that the central and AT-doublet EIT peaks oscillate 180$^\circ$ out of phase as a function of RF phase.

We now turn to the utility of our phase-sensitive quantum interferometric scheme.
As noted above, the relative phases and frequencies of the three applied fields fixes a reference frequency and phase on the fourth transition.
For our measurements in Fig.~\ref{PhaseSweep}(a), we use a frequency locked to that of this reference, which represents a homodyne measurement.
We again emphasize that our reference phase and frequency are not the result of an applied field, but rather are encoded in the quantum mechanical wave functions of the Rydberg states adjacent to our transition. 
This reference can then be used in a heterodyne configuration analogously to a conventional Rydberg mixer \cite{simons2019}. 

\begin{figure*}[!htb]
\centering
\includegraphics[width=.85\linewidth]{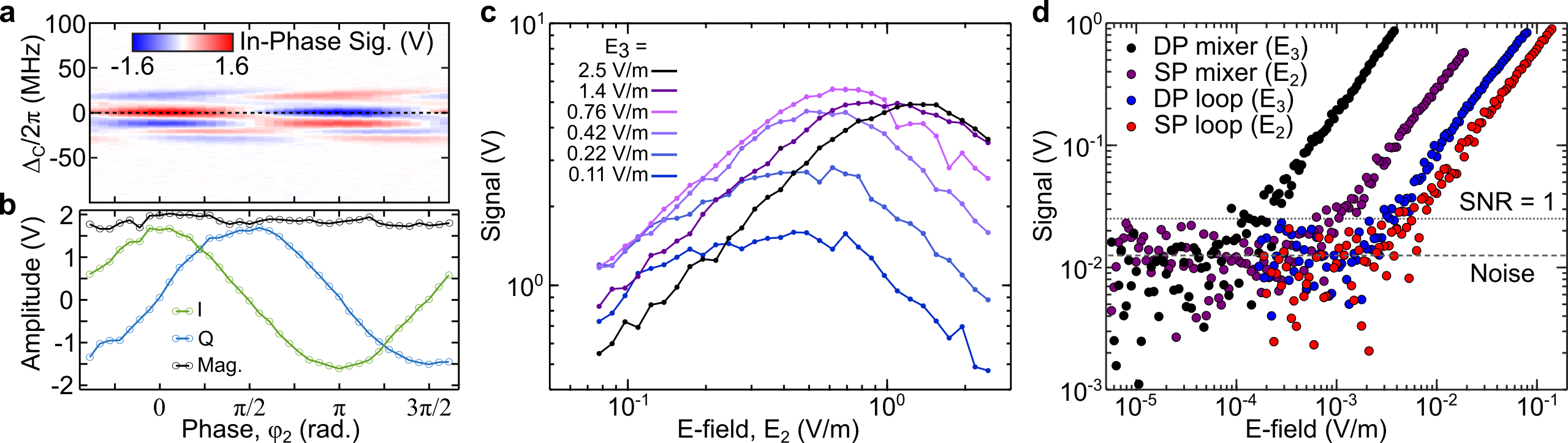}
\caption{\textbf{Quantum Interferometric Rydberg Mixer.}
In-phase (I) lock-in demodulated signal (a) as a function of $\phi_2$, together with (b) a line cut along the signal maximum along with the corresponding quadrature (Q) signal and the lock-in magnitude.
An E-field strength-dependent measurement (c) shows a monotonic mixer sensitivity over a broad range.
Sensitivity measurements (d) compare the field-dependent amplitudes of  Rydberg mixers on the bare DP and SP transitions with sensitivities achieved using the quantum interferometric mixer on the DP and SP transitions independently.
}
\label{mixer}
\end{figure*}

In this approach, one of the RF fields, e.g., $\omega_3$, is applied at a detuned frequency $\omega^\prime_3$~=~$\omega_3$~+~$\delta$.
This detuned frequency is equivalent to a resonant frequency with a time-varying phase, $\omega_3$~+~$\delta$~=~$\omega_3$~+~$d\phi/dt$, and the resulting oscillation in the EIT signal can be demodulated using lock-in detection at frequency $\delta$, where the lock-in phase provides a direct measure of the RF phase.
For Fig.~\ref{mixer}(a)-(c) we detune and demodulate $\omega_3$, which can be used to measure either $\phi_2$ or $\phi_3$.
Shown in Fig.~\ref{mixer}(a) is the $\phi_2$-dependent in-phase (I) lock-in mixer signal as a function of $\Delta_C$, showing the expected 360$^\circ$ phase sensitivity.
The corresponding $\phi_2$-dependent evolution of the in-phase and out-of-phase (quadrature, Q) signals is shown in Fig.~\ref{mixer}(b), taken along the spectral position of the dashed line in (a). 
We can clearly demodulate the I and Q components of the RF signal simultaneously while the overall signal magnitude remains constant. 

Next we examine the dependence of our quantum interferometric Rydberg mixer on electric field strength. 
Shown in Fig.~\ref{mixer}(c) is the mixer magnitude as a function of E$_2$ at different values of E$_3$ as indicated. 
We can clearly see that for all values of E$_3$ there is a corresponding optimum value of E$_2$ that provides maximum sensitivity, where a weaker E$_2$ favors a weaker E$_3$ and vice versa. 
Since the signal magnitude remains monotonic over a large range of E$_2$ this approach can enable both phase and amplitude sensing.

We examine the low-field sensitivity of our approach in Fig.~\ref{mixer}(d).
Here we use a lock-in filter bandwidth of 1~Hz to measure the signal amplitude at low-field strengths for normal Rydberg mixers applied on the SP and DP transitions individually \cite{gordon2019}, as well as our interferometric loop mixer probing the SP and DP resonant transitions. 
The normal Rydberg mixers are measured using an LO field and signal field sourced from two different signal generators to produce a beat note at 10~kHz while the coupling laser Rabi frequencies are idential to those in the loop measurements.
In all cases, the LO or complementary loop RF fields were empirically optimized for low-field sensitivity.
As we are measuring the magnitude (R) output of our lock-in, the noise floor is calculated based on the average of the measured data points below the sensitivity threshold for the DP loop and used to determine the value for a signal-to-noise ratio (SNR) of 1.
 From this, we estimate sensitivities of approximately 0.15~$\rm mV/m \cdot \sqrt{\rm{Hz}}$ and 1~$\rm mV/m \cdot \sqrt{\rm{Hz}}$ for the DP and SP mixers, respectively, and sensitivities of 2~$\rm mV/m \cdot \sqrt{\rm{Hz}}$ and 6~$\rm mV/m \cdot \sqrt{\rm{Hz}}$ for the DP and SP interferometric loop mixers, respectively.

\begin{figure*}[!htb]
\centering
\includegraphics[width=.95\linewidth]{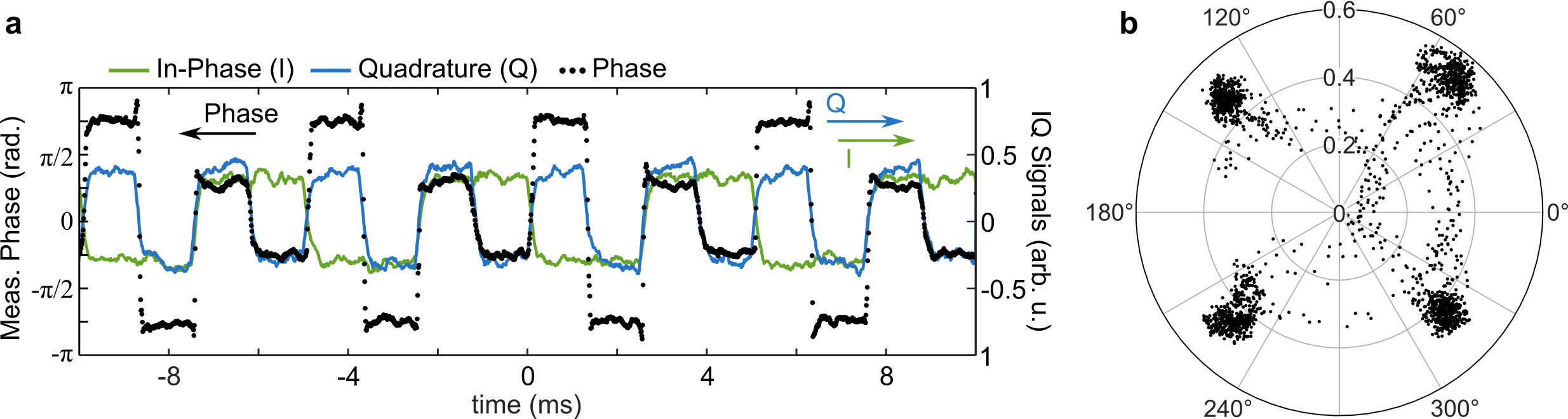}
\caption{\textbf{Signal Demodulation}
The phase plot showing detection of a four phase-state signal using the the quantum phase mixer, along with the corresponding I and Q signals is shown in (a) at a symbol rate of 800 Hz.
The corresponding constellation diagram (b) shows the well-resolved phase-states.
}
\label{qpsk}
\end{figure*}

\section{Discussion}

To demonstrate the general utility of our quantum interferometric loop Rydberg mixer we broadcast a four phase-state $\phi_3$ signal onto the atoms, generated using an IQ mixer to simulate a QPSK signal.
Shown in Fig.~\ref{qpsk}(a) is the lock-in output of our simulated QPSK signal with a symbol rate of 800 Hz, showing the phase (black dots) together with the corresponding orthogonal I and Q channels (green and blue, respectively).
The received constellation diagram in Fig.~\ref{qpsk}(b) shows the four well-resolved phase states detected using our scheme.
\textcolor{black}{We emphasize that our 800 Hz bandwidth was not bandwidth-limited, but without the vector signal generator and analyzer used in previous work \cite{holloway2019a} it was chosen for illustrative purposes based on instrumental and signal-level limitations of our approach}.

The use of our phase-coherent quantum interferometric scheme holds notable advantages compared to conventional Rydberg mixer approaches. 
First off, due to the dependence of the signal on the accumulated phase over the interferometric loop, we can now detune one field to generate a mixer that measures the phase of a different field.
With the necessary presence of at least four fields to complete the loop, this may enable new modulation and frequency mixing schemes for detection and demodulation of RF fields.
The use of degenerate RF frequencies \cite{anderson2022} is a special case of our approach.
Using two degenerate transitions, the RF phase is accumulated on both transitions $\phi_{tot}=\phi_2+\phi_3=2\phi_{RF}$, where the doubled phase provides only 180$^\circ$ phase resolution and thus renders common phase modulation schemes unusable.
And although our present implementation does not do away with RF fields altogether, the ability to apply a field on one RF transition in order to measure another allows frequency separation of the two fields, potentially into distinct bands. 
\textcolor{black}{This may be useful for sensing applications where broadcasting an LO field in the spectral vicinity of the signal of interest is undesirable.}

Our sensitivity measurements show a diminished sensitivity of our loop approach compared to standard Rydberg mixers on the same transitions.
The DP interferometric loop mixer shows an order of magnitude decrease in sensitivity compared to the DP mixer, though it is only a factor of 5 less than the SP mixer. 
This underscores a key conclusion of Fig.~\ref{PhaseSweep}(b)-(d): The overall sensitivity of this approach is maximized when $\Omega_1 = \Omega_4$, i.e., the EIT amplitudes are the same, and we see that it is ultimately limited by the weaker of the two. 
As such, we expect that high sensitivity could nevertheless be achieved with our approach given a better matched state manifold.
\textcolor{black}{Although the nD-(n-1)F-(n+1)D state manifold provides better matched and large transition dipole moments, we have found that the transitions connecting the nD-(n+2)P-(n+1)D states are within a few 100 MHz of the D-F ones and thus adversely affect the sensitivity of the associated loop.
We do expect improvement in Cs, however, where the D-P-D transitions are well-isolated.}

Our overall mixer sensitivity is also reduced by a factor of around 20 relative to record values reported in the literature \cite{prajapati2021,jing2020}.
The E-field sensitivity is a trade-off between high RF transition dipole moments at high $n$ and better state isolation and higher coupling laser transition dipole moments at low $n$.
Thus, it is not surprising that these high sensitivity values are achieved using higher angular momentum states, but they also rely on lower principal quantum numbers in the vicinity of $n\approx50$.
As such, we would expect dramatic improvement using an EOM with higher operating frequency.
\textcolor{black}{In this context it is important to note that although the origin of the discrepancy between our measured S- and D-state EIT amplitudes and those predicted by our model is unclear, this further underscores the benefit of operating at lower values of $n$ where this discrepancy disappears and smaller EIT ratios are seen.}
Concluding that the overall sensitivity of our quantum interferometric scheme can approach that of a traditional LO-based mixer, it is also clear that the sensitivity improvements afforded by an LO-based approach also apply here.

Although we have used a four-photon scheme involving two optical and two RF fields, our demonstration of weak-field sensitivity suggests that it should be possible to replace one of the RF fields with an additional optical one -- whose Rabi frequencies are typically low -- to enable all-optical Rydberg atom E-field sensing.
\textcolor{black}{Eliminating the need for an external RF field altogether is attractive for angle-of-arrival measurements where a phase front (propagation direction) mismatch between the external field and the signal of interest will cause a reduced signal amplitude and phase accuracy.
The inherent challenges of phase-locking three optical fields to obtain a phase-stable loop can be achieved by several possible means.
While potentially difficult to achieve, parametric generation generates phase-locked fields that will cancel any phase noise in the pump field through the loop, while phase-locking using a frequency comb requires three separate lasers but may allow for a broader range of frequencies.
}


\section{Conclusion}

We have demonstrated a Rydberg atom-based RF electric field sensor scheme that uses quantum interference over a closed loop of atomic transitions with phase and frequency fixed by the fields used.
This approach provides full 360$^\circ$ phase-resolved field sensing without an applied local oscillator near the detected frequency. 
We further show that this approach enables LO-free functionality analogous to a traditional LO-based Rydberg mixer, where we demonstrate phase-resolved demodulation of a four phase-state QPSK signal using our quantum interferometric mixer.
These experiments demonstrate the clear-cut advantages of closed-loop quantum interferometric schemes for Rydberg atom-based RF field sensing, and further hold potential for all-optical field sensing.
\\

\section*{Acknowledgements}
This research was developed with funding from the Defense Advanced Research Projects Agency (DARPA).
The views, opinions and/or findings expressed are those of the author and should not be interpreted as representing the official views or policies of the Department of Defense or the U.S. Government.
A contribution of the US government, not subject to copyright in the United States.

\bibliography{quantum_phase_prappl_rev}

\end{document}